\documentstyle[aps,amssymb,12pt]{revtex}
\begin{document}
\title{Pauli principle and chaos in a magnetized disk}
\author{R. Badrinarayanan$^1$, A. G\'ongora-T$^2$, and Jorge V.
Jos\'{e}$^{1}$}
\address{$^{1}$Physics Department and Center for Interdisciplinary
Research on Complex Systems, \\ Northeastern University, Boston,
MA~02115, USA\\
$^{2}$ Centro de Ciencias F\'\i sicas, Universidad Nacional Aut\'onoma
de M\'exico\\ Apartado Postal 48-3, 62250 Cuernavaca, Morelos, MEXICO}
\maketitle
\begin{abstract}
We present results of  a detailed quantum mechanical study of a gas of 
$N$ noninteracting electrons confined to a  circular boundary and 
subject to homogeneous dc plus ac magnetic fields $(B=B_{dc}+B_{ac}f(t)$, 
with  $f(t+2\pi/\omega_0)=f(t)$).  We earlier found a one-particle 
{\it classical} phase diagram of  the (scaled) Larmor frequency 
$\tilde\omega_c=omega_c/\omega_0$ {\rm vs} $\epsilon=B_{ac}/B_{dc}$ that 
separates  regular from chaotic regimes. We also showed that the 
quantum spectrum statistics changed from Poisson to Gaussian orthogonal 
ensembles in the transition from classically integrable to chaotic 
dynamics. Here we find that, as a  function of  $N$ and
$(\epsilon,\tilde\omega_c)$, there are clear quantum signatures 
in the magnetic response, when going from the single-particle 
classically regular to chaotic regimes. In the quasi-integrable 
regime the magnetization  non-monotonically oscillates between 
diamagnetic and paramagnetic as a function of $N$.  We quantitatively 
understand this behavior from a perturbation theory analysis. 
In the chaotic regime, however, we find that the magnetization 
oscillates as a function of $N$ but it  is {\it always} diamagnetic.  
Equivalent results are also presented for the orbital currents. We also 
find that the time-averaged energy grows like $N^2$ in the quasi-integrable 
regime but changes to a linear $N$ dependence in the chaotic regime. 
In contrast, the results with Bose statistics are akin to the 
single-particle case and thus different from the fermionic
case. We also give an estimate of possible experimental parameters
were our results may be seen in semiconductor quantum dot billiards.
\end{abstract}
{Pacs 05.45.+b, 03.65.-w, 72.20.Ht}
\maketitle
%
\newpage
\section{Introduction}
\label{sec:intro}
There is a long history of studies of the magnetic response of an electron gas,
confined to a finite boundary.  Starting with  Bohr and van Leeuwen 
\cite{vanvleck}, to Landau's finite diamagnetism in the quantum regime 
\cite{landau,dingle}. The problem is still of current
theoretical an experimental interest \cite{rev}, in particular due to 
the realization that for  most geometries of the confining boundary one 
can find classically chaotic behavior \cite{studies1}-\cite{levy}.
Most previous studies of this problem have assumed that the external 
magnetic field is static and they have  concentrated in calculating 
the static magnetic susceptibility, except for the dynamic magnetic 
field experimental work of Reulet et al. \cite{reulet}.

In an earlier paper \cite{paper1} (referred to as I hereafter), 
we investigated the classical  dynamics and the quantum
signatures of classical chaos, for {\it one electron} confined to a 
circular quantum dot structure. The dot was subjected to uniform  $d.c.$
($B_{dc}$) plus $a.c.$ ($B_{ac}f(t)$), with periodic
$f(t)=f(t+2\pi/\omega_0)$) perpendicular magnetic fields. There, we 
established an approximate phase boundary in the parameter space spanned by 
$(\epsilon= B_{ac}/B_{dc},\tilde{\omega_c}=\omega_c/\omega_0)$ that separates
the classically regular from the chaotic regimes, where 
${\omega_c}$  is the Larmor frequency of the  $d.c.$ field.
The phase diagram shown in Fig. \ref{fig1}, 
which we shall often use in our analysis here,  separates the 
quasi-integrable from chaotic regimes. In I we established clear 
correspondences between the transitions in the classical behavior 
and their corresponding quantum signatures. From the statistical 
properties of the quasienergy spectrum of the one-period evolution operator, 
going from Poisson to Gaussian orthogonal ensemble, to the  semiclassical 
phase space correspondences via the Husimi quasienergy eigenfunction 
distribution functions.

In this paper we present a detailed quantum mechanical study of the zero 
temperature magnetic response of a noninteracting electron 
gas confined to a  circular boundary and subject to the same combination 
of a $d.c.$ plus $a.c.$  magnetic fields. Here we are interested 
in considering the magnetic response of this model for an $N$ electron system
that satisfies the Pauli exclusion principle.
Another basic question, first addressed in this paper, is how does
the transition from regular to chaotic behavior in the classical case,
where the particles are indistinguishable, affect
their fermionic quantum nature.  Most previous studies of the quantum
manifestations of classical chaos have centered on one-particle
problems. Here we only address the important particle-statistics
many-particle problem, and leave for a future study the relevant
effects of electronic interactions. As we show below there
are indeed clear manifestations of the particle statistics, which 
are different if we are in the classically integrable regime from 
those where the system is chaotic.

The organization of the rest of the paper is as follows: In section 
II we briefly  recapitulate the main elements of the single particle model 
studied in I, together with expressions for the matrix elements of the 
operators needed in our analysis. Next we outline our method to calculate 
the matrix elements of  multi-electron operators in a basis of properly 
(anti)symmetrized  eigenfunctions. In  section III we present our main 
results for the magnetization, orbital currents and energy. We  calculated 
both the time evolution of such operators and their time-averages  as a
function $N$ and the parameters  $(\epsilon,\tilde{\omega_c})$. We also 
include a perturbative calculation, fully described in the Appendix, 
that quantitatively explains our numerical results for the magnetization 
in the quasi-integrable weak-field regime.  Finally, in section IV 
we present  a summary of our conclusions, with an estimate of a few  
experimental parameters that may give an idea of the regimes in frequency
and fields where the transition between integrable and chaotic regimes
discussed in this paper could be tested.

\section{The Model}
\label{sec:model}

\subsection{One-Electron Wavefunction}

We start by recalling the main features of the single particle formalism, 
as explained in paper  I, and next its extension to the N non-interacting
electron problem. The model we consider here is that of electrons
confined to a disk, and subject to a steady ($B_{dc}$)
and a time-periodic  ($B_{ac}$) magnetic field. 
After scaling to appropriate dimensionless units, the model 
Hamiltonian considered  here is, 
\begin{equation}
\label{eq:5-1a}
\tilde H = \tilde H_{dc} + \tilde H_1(\tau), 
\end{equation}
which in polar coordinates reads,
\begin{equation}
\label{eq:5-1b}
\tilde H_{dc} = -\frac{{\tilde\hbar}^2}{2}
\left( \frac{d^2}{dr^2} + \frac{1}{r}\frac{d}{dr}
\right) + \frac{\ell^2 {\tilde\hbar}^2}{2 r^2} + \frac{1}{2}
\left(\frac{{\tilde\omega_c}}{2}\right)^2 r^2 + 
\ell\, {\tilde\hbar} \frac{{\tilde\omega_c}}{2} ,
\end{equation}
and with the time-dependent kick component
\begin{equation}
\label{eq:5-ic}
\tilde H_1(\tau) = \frac{1}{2}\ \eta\ r^2 
\sum_{n=-\infty}^{\infty} \delta (\tau-n). 
\end{equation}
The dimensionless units are defined as, 
\begin{eqnarray}
\label{5-all2}
\label{eq:5-2a}
r = \frac{\rho}{R_0}, \quad 0\le &r& \le 1;\quad 
\tau = \frac{t}{T_0} \equiv \frac{\omega_0}{2\pi}\,t, \\
\label{eq:5-2b}
{\tilde\omega_c} = \frac{\omega_c}{\omega_0}, \quad
{\tilde\hbar} = \frac{\hbar}{m^*\omega_0 R_0^2}, \quad \epsilon &=& 
\frac{B_{ac}}{B_{dc}} = \frac{\omega_{ac}}{\omega_c},  
\quad{\rm and}\quad\eta = \left(\frac{\epsilon\,
{\tilde\omega_c}}{2}\right)^2. 
\end{eqnarray} 
Here $R_0$ is the radius of the disk quantum dot assumed to have 
rigid walls. $T_0$ is the drive period of the $a.c.$ field, 
$\omega_c=e^*B_{dc}/(m^*c)$ is the static Larmor frequency, in 
terms of the effective electron mass $m^*$ $ (\sim 0.067m_e)$,
the screened  electronic charge $e^* $ $(\sim 0.3e)$ \cite{benaaker}, and the
dynamic frequency $\omega_{ac}=e^*B_{ac}/(m^*c)$.  

The exact eigenfunctions of the static Hamiltonian 
$\tilde H_{dc}$ are given in terms of the Whittaker $M$
functions  \cite{dingle},
\begin{equation}
\label{eq:5-a}
\tilde\psi_{n\ell}(r) = \sqrt{\frac{2}{N_{n\ell}}}\; 
\frac{1}{r}\;M_{\chi_{n\ell},{\mid \ell\mid}/2}(\frac{f}{2} r^2),
\end{equation}
with $n$ the principal quantum number, $\ell $ the angular 
momentum eigenvalue, and $N_{n\ell}$ a normalization constant. 
The frustration parameter, $f$, that measures the number of
flux quanta in the disk,  is defined by,
 
\begin{equation}
\label{eq:5-b}
f = \frac{\Phi}{\Phi_0} \equiv 
\frac{B_{dc}\pi R_0^2}{\left(hc/2e^*\right)} \equiv  
\frac{\tilde{\omega_c}}{\tilde\hbar}=(\frac{R_0}{\ell _B})^2, 
\end{equation}
with $\ell_B=(\frac{\hbar c}{eB_{dc}})^{1/2}$ the magnetic 
length and $\Phi_0 = \frac{hc}{2e^*}$  the quantum of flux.   
The eigenenergies   
\begin{equation}
\label{eq:5-c}
\tilde E_{n\ell} = 2 ( \chi_{n\ell} + \ell ),
\end{equation}
are determined  by the requirement that the wavefunction 
vanishes at the boundary i.e.,  by the zeros  of the Whittaker function,
$M_{\chi_{n\ell},{\mid \ell\mid}/2}(\frac{f}{2}) = 0$.

We calculated the energy eigenvalues $\tilde E_{n\ell}$ for the static 
problem in a basis  of Whittaker functions  as a function of $B_{dc}$, 
and checked our numbers by fully reproducing the results of
Ref.\cite{studies1}.  The Whittaker functions  have the advantage
of being valid over the entire 
range of parameters, however, they are numerically difficult to evaluate 
for the full time-dependent problem. For convenience
when calculating the time-dependent problem, 
we decided also to expand the total (single particle) wavefunction 
in a Fourier Sine-basis. In this case
\begin{eqnarray}
\label{5-all3}
\label{eq:5-3a}
\langle r |\tilde H_{dc}\, |\tilde\psi_{n\ell}(\phi)\rangle &=& 
\tilde E_{n\ell}\, 
\tilde\psi_{n\ell}(r)  \frac{e^{i \ell \phi}}{\sqrt{2 \pi}},\\
\label{eq:5-3b}
\langle r|\tilde\Psi(\phi)\rangle &=& \sum_{n=1}^{\infty} 
\sum_{\ell=-\infty}^{\infty} 
\tilde\psi_{n\ell}(r) \frac{e^{i \ell \phi}}
{\sqrt{2 \pi}},\\
\label{eq:5-3c}
\tilde\psi_{n\ell}(r = 1) &=& 0, \qquad \qquad 
\int_{0}^{1} \tilde\psi_{n\ell}^{2}(r) \, r \,dr = 1,\\
\label{eq:5-3d}
{\rm and} \,\,\,
\langle r|\tilde \psi_{n\ell}\rangle &=& \sqrt{\frac{2}{r}} 
\sin(n\pi r).
\end{eqnarray}

This basis set is properly orthonormalized, and automatically 
satisfies the boundary conditions. To calculate the spectrum 
of the static problem, we used, nonetheless,  the exact eigenvalues 
of $\tilde H_{dc}$, given by the zeros of the  Whittaker functions. 
Doing this allowed us also to check the reliability of our Sine-basis 
numerical method. 

We then computed the required matrix elements of the operators 
we are interested in, within the Sine-basis method.
For example, for the magnetization operator 
\begin{equation}
\label{eq:5-4}
\mbox{\boldmath{$\mu$}} = \frac{e^*}{2m^*c}
\left( \mbox{\boldmath{$\cal L$}} 
- \frac{e^*}{c}~\mbox{\boldmath{r}}\times
{\rm{\bf A}}(\mbox{\boldmath{r}})\right),
\end{equation}
where \mbox{\boldmath{$\cal L$}} is the angular momentum operator and 
${\rm{\bf A}}(\mbox{\boldmath{r}})$ is the electromagnetic
vector potential in normalized coordinates. In the present case, we take the
magnetic field perpendicular to the plane, then the z-component 
of the magnetization operator is
\begin{equation}
\label{eq:5-5}
\tilde M_z(r) = \frac{\hat M_z}{\mu_B} = -\frac{\hat L_z}{\tilde\hbar} - 
\frac{f}{2} \hat r^2,
\end{equation}
with $L_z$ the z-component of the angular momentum,
$\mu_B = \frac{|e^*|\hbar}{2m^*c}$ the Bohr magneton, and $\hat M_z$ 
the magnetization operator along the z-axis.  The matrix elements of  
$\tilde M_z$  in the Fourier Sine-basis are given by,
\begin{eqnarray}
\label{alleq5-6}
\langle m|\tilde M_z|n\rangle = &-&\Big\{\ell 
+ \frac{f}{2}\left(\frac{1}{3}-\frac{1}
{2n^2\pi^2}\right) \Big\}~\delta_{mn} \nonumber \\
&-&\Big\{ \frac{f}{2} ~\frac{(-)^{m+n}}{\pi^2}\frac{8mn}
{(m^2-n^2)^2} \Big\}~(1-\delta_{mn}).
\end{eqnarray}
Similarly, starting from the definition of the current density operator  
\begin{equation}
\label{eq:5-7}
{\rm {\bf J}} = \frac{1}{2m^*}\left(-i\hbar
\mbox{\boldmath{$\nabla$}} - 
\frac{e^*}{c}{\rm{\bf A}}\right) + c.c.,
\end{equation}
(where $c.c.$ stands for  complex conjugate), 
we have the following expression for the azimuthal current 
densities: $J_{\phi} = J_{\phi}^{(para)} + J_{\phi}^{(dia)}$, where   
\begin{equation}
\label{eq:5-8}
J_{\phi}^{(para)} = -\frac{i\tilde \hbar}{2} 
\frac{1}{r}\frac{\partial}{\partial\phi}
+ c.c. \quad {\rm and}, \quad 
J_{\phi}^{(dia)} = \frac{\tilde \omega_c}{2} r,
\end{equation}
are the paramagnetic and diamagnetic current densities, respectively. 
In the Fourier Sine-basis, the matrix elements of the 
current densities 
(in units of $\tilde \hbar$) are given by
\begin{eqnarray}
\label{alleq5-9a}
\label{eq:5-9}
\langle m|J_{\phi}^{(para)}|n\rangle &=& 
\ell{\tilde \hbar}\cdot
\left\{\begin{array}{ll}
      - {\rm Ci}[2n\pi] + \gamma_E + {\rm ln}(2n\pi) & \mbox{$(m=n)$} \\
      - {\rm Ci}\left[(m+n)\pi\right] + {\rm Ci}\left[(m-n)\pi\right] & 
        \mbox{$(m\ne n),$}\end{array} 
\right. \\
\label{eq:5-9b}
\langle m|J_{\phi}^{(dia)}|n\rangle &=& 
\frac{f}{2}{\tilde \hbar}\cdot
\left\{\begin{array}{ll}
       \frac{1}{2} & \mbox{$(m=n)$} \\
       \frac{(-)^{m+n}-1}{\pi^2} \frac{4mn}{(m^2-n^2)^2} & 
       \mbox{$(m\ne n)$}\end{array}
\right.
\end{eqnarray}
where $\gamma_E = 0.57721~566649\dots$ is Euler's gamma number, 
and $\rm{Ci}(x)$ is the Cosine integral.

Finally, the expression for the one-period time-evolution
operator $U_{\ell}(\tau,\tau_0)$,
for the single particle Hamiltonian, that satisfies the
 dynamical equation  
\begin{equation}
\label{eq:5-10}
i {\tilde\hbar}\,\frac{\partial}{\partial\tau}\,U_{\ell}
(\tau,\tau_0) = ( {\tilde H_{dc}} + {\tilde H_1(\tau)} )\,
U_{\ell}(\tau,\tau_0),
\end{equation}
is
\begin{equation}
\label{eq:5-11}
U_\ell(1,0) = \exp\left(-\frac{i}{{\tilde\hbar}}\,\frac{1}{2}
\eta\,r^2\right)\,\exp\left(-\frac{i}{{\tilde\hbar}}\,
{\tilde H_{dc}}\right) .
\end{equation}
The total (single particle) wavefunction at any integer multiple 
$N_T$ of the period (hereafter taken to be 1), is given by 
repeated applications of $U_{\ell}$ to the initial wavefunction:
\begin{equation}
\label{eq:5-12}
|\Psi_{\ell}(r,\phi,N_T)\rangle = U_{\ell}^{N_T}\,
|\Psi_{\ell}(r,\phi,0)\rangle .
\end{equation}

\subsection{Many-Electron Wavefunctions}

One can directly generalize the above single-electron 
formalism to the many-electron case. Take the initial $N$-electron 
wavefunction to be
\begin{equation}
\label{eq:5-13}
|\Phi(\mbox{\boldmath{$r$}}_1,
\mbox{\boldmath{$r$}}_2\dots ,
\mbox{\boldmath{$r$}}_N)\rangle \equiv 
|\Phi(1,2\dots ,N)\rangle,
\end{equation}
which is antisymmetric under exchange of an odd number of 
particles (the Pauli exclusion principle) :
\begin{equation}
\label{eq:5-14}
|\Phi(1,\dots i,\dots j,\dots ,N)\rangle = 
-|\Phi(1,\dots j,\dots i,\dots ,N)\rangle.
\end{equation}
Let the $i$-th single-particle eigenstate satisfy
the equation
\begin{equation}
\label{eq:5-15}
\tilde H^{(i)}|\Psi_{n_{i}\ell_{i}}(i)\rangle = 
E_{n_{i}\ell_{i}}^{(i)}|\Psi_{n_{i}\ell_{i}}(i)\rangle. 
\end{equation}
We know that the Slater antisymmetrization 
procedure for 
the non-interacting $N$-electron state can be written as 
the following tensor product \cite{huang}: 
\begin{eqnarray}
\label{alleq5-16} 
{\rm{\bf\hat A}}&\cdot&|\Phi(1,2\dots ,N)\rangle 
= \frac{1}{\sqrt{N!}}\left|\begin{array}{llll}
|\Psi_{n_1\ell_1}(1)\rangle & 
\otimes|\Psi_{n_2\ell_2}(1)\rangle & 
\cdots & 
\otimes|\Psi_{n_N\ell_N}(1)\rangle \\
|\Psi_{n_1\ell_1}(2)\rangle & 
\otimes|\Psi_{n_2\ell_2}(2)\rangle & 
\cdots & 
\otimes|\Psi_{n_N\ell_N}(2)\rangle \\
\vdots & 
\vdots & 
\ddots & 
\vdots \\
|\Psi_{n_1\ell_1}(N)\rangle & 
\otimes|\Psi_{n_2\ell_2}(N)\rangle & 
\cdots & 
\otimes|\Psi_{n_N\ell_N}(N)\rangle 
\end{array}\right| \nonumber \\
&=& \frac{1}{\sqrt{N!}}\sum_{P} \delta_{P}~\Bigg[
|\Psi_{P\{n_1\ell_1\}}(1)\rangle\otimes
|\Psi_{P\{n_2\ell_2\}}(2)\rangle\cdots
\otimes|\Psi_{P\{n_N\ell_N\}}(N)\rangle\Bigg]
\end{eqnarray}
where ${\rm {\bf \hat A}}$ is the antisymmetrization operator, and 
$P$ is the permutation operator 
\begin{eqnarray}
\label{5-all17}
\label{eq:5-17a}
P\{1,2,\dots ,N\} &=& \{P1,P2,\dots ,PN\},\\
\label{eq:5-17b}
\delta_{P} &=& \left\{\begin{array}{ll} 
+1 & \mbox{(even $P$)} \\
-1 & \mbox{(odd  $P$)} 
\end{array} \right. .
\end{eqnarray}
The summation runs over all possible permutations. Furthermore, 
the trace of any sum of $N$-body operators
${\hat O} = \sum_{i=1}^N {\hat O_i}$ 
can be written as 
\begin{eqnarray}
\label{alleq5-18}
{\rm Tr}\{{\hat O}\} &=& \langle\Phi(1,\dots ,N)|
{\hat O}|\Phi(1,\dots ,N)\rangle \nonumber \\ 
&=& \frac{1}{N}\sum_{i,j=1}^{N}
\langle\Psi_{n_{i}\ell_{i}}(j)|{\hat O_{j}}
|\Psi_{n_{i}\ell_{i}}(j)\rangle.
\end{eqnarray}

Now consider the time evolution of such a system. 
For simplicity, we take as the initial state the lowest energy 
(ground) state of the {\it unperturbed} system 
({\sl i.e.,} without the $a.c.$ field) allowed 
by the Pauli Principle : 
\begin{equation}
\label{eq:5-19}
|\Phi_{1,\dots, N}(t=0)\rangle = {\rm{\bf\hat{A}}}\cdot\left\{
|\Psi_{n_{1}\ell_{1}}(1)\rangle\otimes 
|\Psi_{n_{2}\ell_{2}}(2)\rangle
\cdots\otimes |\Psi_{n_{N}\ell_{N}}(N)\rangle\right\}.
\end{equation}
The one-period time-evolution operator for the N-electron system
is given by the tensor product,
\begin{equation}
\label{eq:5-20}
\mbox{\boldmath{$U$}} = U_1(1,0)\otimes U_2(1,0)\cdots
\otimes U_N(1,0).
\end{equation}
Since the antisymmetrization and time-evolution operators commute,  
the state after $N_T$ periods is simply given by  
\begin{eqnarray}
\label{eq:5-21}
|\Phi_{1,\dots, N}(N_T)\rangle &=& 
\mbox{\boldmath{$U$}}^{N_T}
\cdot|\Phi_{1,\dots, N}(t=0)\rangle \nonumber \\ 
&=& {\rm{\bf\hat{A}}}\cdot\left\{U_1^{N_T}
|\Psi_{n_1\ell_1}(1)\rangle \cdots\otimes
U_N^{N_T}|\Psi_{n_N\ell_N}(N)\rangle\right\},
\end{eqnarray}
where we've used the notation $U_i(1,0)\equiv U_i,$ for $i=1,\ldots,N$. 
We can now generalize Eq. (\ref{alleq5-18}) for the trace 
of an operator at any integer multiple $N_T$ of the period as, 
\begin{eqnarray}
\label{eq:5-22}
{\rm Tr}\{{\hat O}\}(t=N_T) 
&=& \langle\Phi_{1,\dots, N}(t=N_T)|{\hat O}
|\Phi_{1,\dots, N}(t=N_T)\rangle \nonumber \\
&=& \frac{1}{N}\sum_{i,j=1}^{N}\langle\Psi_{n_{i}\ell_{i}}(j,t=0)|
(U_j^{\dagger})^{N_T}
{\hat O_{j}}U_j^{N_T}|\Psi_{n_{i}\ell_{i}}(j,t=0)\rangle.
\end{eqnarray}
In particular, for the average quantum time-dependent magnetization 
{\it per electron} we have
\begin{equation}
\label{eq:5-23}
\frac{\langle{\tilde M_z}\rangle(N_T)}{N} = 
-\frac{L}{N} + \frac{1}{N^2}
\sum_{i,j=1}^N \langle\Psi_{n_{i}\ell_{i}}(j)|
(U_j^{\dagger})^{N_T}
(-\frac{f}{2} {r}_j^2)U_j^{N_T}|\Psi_{n_{i}\ell_{i}}(j)\rangle,
\end{equation}
where $L=\sum_{i=1}^{N} \ell_i$. Similarly, we can write the corresponding
expressions for the time-dependent orbital currents, and the
total time-dependent Hamiltonian average, which we term the averaged energy.
For example, the {\it time-averaged magnetization}, 
$\langle\langle M_z \rangle\rangle$, is  defined by
\begin{equation}
\label{eq:5-23b}
\langle\langle M_z \rangle\rangle = \lim_{N_T\rightarrow\infty}
 \frac{1}{N_T} \sum_{n=1}^{N_T} \langle M_z \rangle(n).
\end{equation}
\section{Results}
\label{sec:res}

We now come to the discussion of the main results of this paper,
that are concerned with the dynamic and time--averaged properties of 
different relevant operators. The most striking
features are observed in the magnetization of the system, which we 
shall discuss first as a function of the number $N$ of electrons
and ($\epsilon,\tilde\omega_c$).
We also give results for the orbital current as well as 
interesting results for the time-averaged energy as a function
of $N$ in different ($\epsilon,\tilde\omega_c$) parameter regimes.

\subsection{Magnetization, orbital currents and energy}

We start with the time--dependent dynamics of the magnetization 
and its corresponding power spectra for a single electron. 
The power spectrum $S(\nu)$ is the
square of the Fourier transform of the expectation value of the
magnetization operator $\langle\tilde M_z\rangle(t)$. 
(For notational simplicity, we write $\langle\tilde M_z\rangle(t)$ by 
$\langle M\rangle(t)$.) We keep $\tilde\omega_c$ 
fixed, and sweep through values of $\epsilon$, from small to large. 
As can be seen from the phase diagram in Fig. \ref{fig1}, 
for fixed $\tilde \omega_c$, as $\epsilon$ increases the 
underlying classical dynamics changes from quasi-integrable
to chaotic.  

In Fig. \ref{fig2}(a) we see that $\langle M \rangle$ is diamagnetic 
in the regular regime and oscillates periodically with time, reflected 
in the very strong peak in its power spectrum shown in Fig. \ref{fig2}(b).
As we increase the values of $\epsilon$ ({\sl i.e.}, as we approach
the chaos border in the phase diagram), the intermediate dynamics gets
more complex, as shown in Figs. \ref{fig3}(a) and \ref{fig3}(b).
We see in Fig. \ref{fig3}(b), that there are two peaks in $S(\nu)$,
which are due to the quasi-beats seen in Fig. \ref{fig3}(a). In the chaotic
region, $<M>(t)$ shows essentially irregular behavior, Fig. \ref{fig4}(a),
while $S(\nu)$ has a broad background, shown in Fig. \ref{fig4}(b). 
Note that the average value of $<M>(t)$ increases in magnitude, 
becoming steadily more diamagnetic in the chaotic regime.

The one-electron behavior changes significantly with the addition of 
more electrons. The pattern of change from regular to chaotic
is similar as in the one-electron case as we sweep through
the same $\epsilon$-values  as above. The spectral function
develops more resonances in the regular region,
whereas in the chaotic regime it has a broad band spectrum. As we
continue to increase the number of electrons there are more ``beats"
in the time--dependence of the magnetization and more peaks in the
spectral function. 

It is then more convenient to consider the time-averaged properties of 
the magnetization, the orbital current, or the total energy, as a function
of the number of electrons. It is in this type of function that we can 
see important qualitative differences that represent the changes from 
the classical regular to chaotic behavior in the quantum dynamics. 
In Fig. \ref{fig5}(a) we see that for two-electrons the time-averaged 
magnetization in the regular regime is {\it paramagnetic}, whereas for three
electrons becomes diamagnetic again. We note that the dia- to para-magnetic 
changes are non-monotonic  as a function of $N$. For example,
for $N=4$ it switches back to paramagnetic, but remains diamagnetic for
both $N=5$ and $N=6$. A similar situation occurs with the orbital
current as shown in Fig. \ref{fig5}(b).
We mention that the specific value of the frustration parameter ($f =
\frac{{\tilde\omega_c}}{\tilde\hbar}$) determines
if the magnetization flips from dia- to para-magnetic as we
keep adding electrons. Basically, this phenomenon occurs when
$f\sim O(1)$, i.e. when we add  one flux quantum to the dot.
In all other cases, the magnetization remains diamagnetic
and monotonically increasing in magnitude as the number
of electrons increases. We provide a theoretical perturbation 
theory explanation of these dia- to para-magnetic transitions result
in the next subsection.

As we increase the value of $\epsilon$, we enter the chaotic regime.
There we find that for {\it all} electron numbers
the magnetization is {\it always diamagnetic}, at least up to the
maximum number of electrons we considered ($\sim 25)$.
In Fig. \ref{fig5}(c) we show the time-averaged magnetization in
the chaotic regime, which also oscillates as a function of $N$, but it
is always negative and of larger magnitude than in
the quasi-regular regime. A similar situation
occurs for the orbital current (as shown in Fig. \ref{fig5}(d), 
although it has less sharp changes as a function of $N$
than does $<<M>>$.)

In Fig. \ref{fig6}(a) we consider the time-averaged magnetization for a
fixed value of $N=1$, $\epsilon =0.1$ and $\tilde \hbar =0.1$
as a function of $\omega_c$. In this quasi-integrable regime we see
that $<<M>>$ is diamagnetic and decays quadratically as a function of
$\omega_c$. The situation changes in the chaotic regime,
shown in Fig. \ref{fig6}(c), where there is also decay with $\omega_c$
but now the behavior is not as smooth as in the quasi-integrable regime.
We show in Fig. \ref{fig6}(b) the behavior of the time averaged
energy as a function of the number $N$ of electrons. Here we see a clear
quadratic growth  as a function of $N$. 
The situation is remarkably different when the single-particle
classical dynamics is chaotic. In this case, shown in Fig. \ref{fig6}(d), 
the time-averaged energy grows clearly {\it linearly} with $N$. 
This implies that the classically chaotic solutions do have a significant 
quantum signature in the averaged energy, that changes the  quadratic 
quasi-integrable regime behavior to a linear $N$ dependence in the 
chaotic regime. We now present a simple heuristic argument
as to why the  change over between quadratic and linear
$N$ behavior is actually directly related to the Pauli exclusion principle.
We note that in the zero magnetic field case,  each of
the $N$ electrons in the circular dot of radius $R_0$ 
occupies an exclusion principle space  of order $R_0/N^{1/2}$,
while the static free particle kinetic energy changes like $N/R_0^2$.
We expect that this situation does not change much when we are
in the quasi-integrable regime, for finite fields and low frequencies. 
In the classically chaotic regime, in the presence of stronger magnetic 
fields or higher frequencies, the magnetic field will tend,  on the average,
to localize more  the electrons to Larmor orbits inside the dot and in the 
boundaries.  When the field is larger, so that the Larmor radius and $R_0$ 
are comparable, the Landau levels have to be taken into account. In this 
case the electrons will not necessarily feel the presence of the boundary 
and they will remain localized in their ``chaotic'' Landau orbits due to the
time-dependent kicks. In this limit the contribution from
the kinetic energy is much less relevant, and the Larmor orbit radius
will be less dependent of $N$ and $R_0$.

\subsection{Perturbative evaluation of the magnetization in the
quasiregular regime}

In this subsection we present a perturbative analysis that provides
an explanation for the magnetization oscillations as a function
of $N$ in the quasi-integrable regime. Let us first consider the
time-independent part of the one-electron Hamiltonian ${\tilde H}$ , 
and write it as,
\begin{equation}
\label{eq:5-24}
\tilde H_{dc} = \tilde H_{0} + \tilde V, 
\end{equation}
where,
\begin{equation}
\label{eq:5-25}
\tilde H_{0} = -\frac{{\tilde\hbar}^2}{2}
\left( \frac{d^2}{dr^2} + \frac{1}{r}\frac{d}{dr}
\right) + \frac{\ell^2 {\tilde\hbar}^2}{2 r^2} + 
\ell\, {\tilde\hbar} \frac{{\tilde\omega_c}}{2} ,
\end{equation}
and,
\begin{equation}
\label{eq:5-26}
{\tilde V} = \frac{1}{2}\left(\frac{{\tilde\omega_c}}{2}\right)^2 r^2.
\end{equation}
We will consider the limiting case of very small $B_{dc}$ field, {\sl i.e.}, 
${\tilde\omega_c}\ll 1$. As was first shown by Dingle, 
one can write the eigenvalues to first order in ${\tilde\omega_c}$, by 
considering the zero field basis functions of the disk Bessel eigenfunctions:
\begin{equation}
\label{eq:5-27}
{\tilde H_0} |\psi_{n\ell}^{(1)}\rangle \simeq E_{n\ell}^{(1)} 
|\psi_{n\ell}^{(1)}\rangle , 
\end{equation}
where the normalized eigenvalues are \cite{dingle}, 
\begin{equation}
\label{eq:5-28}
E_{n\ell}^{(1)} = \frac{\alpha_{n\ell}^2}{2f} + \ell + \frac{f}{12}
\left\{1 + \frac{2(\ell^2-1)}{\alpha_{n\ell}^2} \right\} + 
O({\tilde\omega_c}^2). 
\end{equation}
Here $\alpha_{n\ell}$ is the $n$th zero of the Bessel function 
$J_{\ell}(x)$, 
and the {\it unperturbed} basis functions are given by
(we consider only the  radial part, the angular part is clear) 
\begin{equation}
\label{eq:5-29}
\langle r|\psi_{n\ell}^{(0)}\rangle = 
\frac{\sqrt{2}}{J_{\ell+1}(\alpha_{n\ell})}
\,J_{\ell}\left(\alpha_{n\ell}r\right) .
\end{equation}
The  matrix elements of the perturbation are then given by,
\begin{equation}
\label{eq:5-30}
\langle\psi_{m\ell}^{(0)}|{\tilde V}|\psi_{n\ell}^{(0)}\rangle \equiv 
{\tilde V_{mn}} = \frac{1}{2}\left(\frac{{\tilde\omega_c}}{2}\right)^2
\left\{\begin{array}{cc} \frac{1}{3}\left[1+\frac{2(\ell^2-1)}
{\alpha_{n\ell}^2}\right] & m=n  \\
\strut{
\frac{8\alpha_{m\ell}\alpha_{n\ell}}{(\alpha_{m\ell}^2
-\alpha_{n\ell}^2)^2} } & m\ne n \end{array} \right.
\end{equation}
The perturbed non degenerate eigen-functions are obtained from
standard perturbation analysis, 
\begin{eqnarray}
\label{all5-31}
|\psi_{n\ell}^{(1)}\rangle &=& |\psi_{n\ell}^{(0)}\rangle + \sum_{m\ne n}
\frac{{\tilde V_{mn}}}{E_{m\ell}^{(0)}-E_{n\ell}^{(0)}} 
\,|\psi_{m\ell}^{(0)}\rangle 
+ O\left({\tilde V_{mn}}^2\right) \nonumber \\ 
&=& |\psi_{n\ell}^{(0)}\rangle + 2f{\tilde\omega_c}^2
\sum_{m\ne n} \frac{\alpha_{m\ell}\alpha_{n\ell}}
{(\alpha_{m\ell}^2-\alpha_{n\ell}^2)^3}|\psi_{m\ell}^{(0)}\rangle 
+ O\left(\alpha^{-8}\right),
\end{eqnarray}
where we've used the unperturbed energy levels 
$E_{m\ell}^{(0)}-E_{n\ell}^{(0)} = 
\frac{1}{4}(\alpha_{m\ell}^2-\alpha_{n\ell}^2)$.
Using Eq.(\ref{all5-31}), and  the definition of the
magnetization operator, Eq.(\ref{eq:5-5}), the 
leading first order matrix element contribution to ${\tilde M_z}$ is given by
\begin{eqnarray}
\label{all5-32}
\langle\psi_{n\ell}^{(1)}|{\tilde M_z}|\psi_{n\ell}^{(1)}\rangle \equiv 
{\langle{\tilde M_z}\rangle}_{mn} &=& -\ell - 2f 
\langle\psi_{n\ell}^{(0)}|r^2|\psi_{n\ell}^{(0)}\rangle + 
O({\tilde\omega_c}^2) \\
&=& -\ell - \frac{2f}{3}\left[1+\frac{2(\ell^2-1)}
{\alpha_{n\ell}^2}\right] + O({\tilde\omega_c}^2) .
\end{eqnarray}
Once again, if we take as the initial state the lowest
energy state allowed  by the Pauli Principle, 
\begin{equation}
\label{eq:5-33}
|\Psi_{1,\dots ,N}(0)\rangle = {\rm{\bf\hat{A}}}\cdot\left\{
|\psi_{n_{1}\ell_{1}}(1)\rangle\otimes 
|\psi_{n_{2}\ell_{2}}(2)\rangle
\cdots\otimes |\psi_{n_{N}\ell_{N}}(N)\rangle\right\}, 
\end{equation}
we can generalize Eq.(\ref{all5-32}) to the $N$-electron case. We 
find that the averaged magnetization {\it per electron}, to
first order approximation, is 
\begin{equation}
\label{eq:5-34}
\frac{\langle{\tilde M_z}\rangle}{N} = -\frac{2f}{3} - \frac{1}{N}
\sum_{i=1}^N \left\{\ell_i + \frac{4f}{3}\,\frac{(\ell_i^2-1)}
{\alpha_{n_i\ell_i}^2} \right\} + O({\tilde\omega_c}^2) .
\end{equation}

Next, we perform a linear-response theory analysis of the {\it full} 
time-dependent problem, assuming that $\epsilon\ll 1$.
We show in the Appendix that within the perturbative approximation
for a single electron, the average magnetization at time
$N_T$ is given by, 
\begin{eqnarray}
\label{eq:eq5-35}
\langle{\tilde M_z}\rangle(N_T) &\simeq& -\left\{\ell_i + 
\frac{f}{2}\langle r^2\rangle_{n_i,n_i}\right\} \nonumber \\
&+& \frac{f}{{2\tilde\hbar}}\left(
\frac{\epsilon{\tilde\omega_c}}{2}\right)^2\sum_{p=1}^{N_T}\sum_{n\ne n_i}
\langle r^2\rangle_{n,n_i}^2 \sin\left\{\omega_{n_i,n}(N_T - p)\right\} , 
\end{eqnarray}
where 
\begin{equation}
\label{eq:5-36}
\langle r^2\rangle_{n,n_i} = \frac{1}{3}\left[1+\frac{2(\ell_i^2-1)}
{\alpha_{n_i\ell_i}^2}\right] + O({\tilde\omega_c}^2) .
\end{equation}
Clearly, the last term in Eq.(\ref{eq:eq5-35}) is of 
$O({\tilde\omega_c}^3)$, so it can be ignored within the current
approximation, which means that to lowest order, the time-dependence
plays no significant role in determining the average magnetization. 

To test the approximation for $\langle{\tilde M_z}\rangle/N$, 
Eq.(\ref{eq:5-34}), we show in Fig. \ref{fig7}, a comparison of 
results from the perturbative and numerically exact calculations. 
The perturbative results agree remarkably well with the 
numerical calculations. We can now understand why the
averaged magnetization oscillates in sign in the regular regime, and it is
because $\langle{\tilde M_z}\rangle/N$ depends most strongly 
on $\sum_{i=1}^N \ell_i$. Whenever this sum of angular momentum 
quantum numbers flips sign, so does the magnetization. For example, for 
the parameters shown in Fig. \ref{fig7}, the values of $F(N) = \sum_{i=1}^N \ell_i$ for 
successive values of $N$ are 
\begin{equation}
\label{eq:5-37}
F(N=1,2,\ldots,12) = 0,-1,0,-2,0,0,-3,0,-1,0,-4,0, 
\end{equation}
corresponding to 
\begin{equation}
\ell_i = (0,-1,1,-2,2,0,-3,3,-1,1,-4,4),
\end{equation} 
i.e., the $\ell$-values of the twelve-electron (unperturbed) ground state.
This is a selection rule associated with the symmetries present in 
the system. 
 
What is interesting is that in the chaotic region, there is no
such flipping of the sign. This difference may constitute an
experimentally accessible signature of chaos in the
quantum system. Clearly, such behavior is 
exclusively a consequence of the Pauli principle, for we would 
not observe any change in the response per electron without it,
since ignoring it in the multi-electron case 
would lead to a trivial rescaling of the single electron results.

\section{Conclusions}
\label{sec:conc}

To summarize, we have studied a model of a non-interacting $N$-electron 
system,  confined to a circular structure with rigid boundaries, 
and subjected to perpendicular constant and time-periodic magnetic fields. 
We studied  the magnetization and orbital currents as a function of 
time, as well as the time-averaged  magnetization 
$\langle\langle {\tilde M_z}\rangle\rangle$, and energy as a function  
of electron number $N$.  We can make a strong connection between the 
dynamic response of  $\langle{\tilde M_z}\rangle(t)$ (or it's power 
spectrum), and the underlying  classical dynamics -- as the classical 
system makes a transition to chaos as we vary the applied magnetic fields, 
the dynamics changes from being harmonic to essentially noisy. There are 
three central significant conclusions: first, the Pauli Principle affects 
the behavior of this non-interacting system significantly, e.g. in 
terms of oscillations of $\langle\langle {\tilde M_z}\rangle\rangle$ as 
a function of $N$. This behavior is directly related to the Pauli 
principle that allows the electrons to optimally reduce their averaged 
$\langle\langle M\rangle\rangle$ at specific values of the total angular 
momentum. Second, while these oscillations in the quasi-integrable 
regime cause the system to flip back and forth between dia- and 
para-magnetic behavior, the system remains diamagnetic at all times 
in the chaotic regime. We also found a very interesting change in the 
time-averaged energy as a function of $N$, going from quadratic in 
the quasi-integrable regime to linear in the chaotic one. We provided a 
simple heuristic explanation of this behavior related to Pauli's
exclusion principle.

In this paper we have not considered the effects of Coulomb interactions that
can significantly complicate the analyses. There are static studies that
have considered the changes in the classical dynamics due to 
interactions. What has been found in some examples is that
if the system of non-interacting particles was non-chaotic, as the
interaction parameter increases the dynamics can become chaotic 
\cite{inter1}. In the quantum
regime the Random Matrix Theory 
statistics can exhibit a transition from
Poisson to  orthogonal ensemble  
as the interaction strength increases \cite{inter2}. What happens in the
time-dependent case considered in this paper that deals with quasi-energy
statistics is not known at present. 

Here we have considered a circular disk in the presence of a
time dependent magnetic field. It is only when we have the ac component
of the field added to the dc one 
that chaos appears. In contrast, if the field is static
but the geometry is changed one can have chaotic classical solutions.
The relevance of the Pauli principle as seen in the zero temperature
magnetization has been studied, for example, by \cite{studies1}. At 
present we do
not know what happens when 
the geometry is not circular and we have a gas of Pauli electrons in the
presence of a dc+ac magnetic field. We expect to consider the two  problems
mentioned above in the future.

To conclude, we briefly give some estimates in terms of physical 
units of the field  strengths and frequencies required to observe 
the effects predicted by our model calculations. In a GaAs-AlGaAs 
semiconductor the radius $R_0$ of a quantum dot device  
\cite{marcus1},\cite{levy}  can be between 0.1 and 10$\mu$m, a sheet 
electron density $n\sim 10^{11}$ cm$^{-2}$, a mobility 
$\mu\sim 265\,000$ cm$^2$/V$\cdot$s, and a characteristic
level spacing  $\Delta\epsilon \sim 0.05$ meV or $\sim 0.5$ K.
In the ballistic electronic motion regime  the elastic mean free path 
$\l_\phi\sim 10\mu$m, with phase coherence length varying 
between 15 and 50 $\mu$m. Typically the power injected  is 
smaller than $1$ nW, which is necessary to avoid electron heating.
For a dot radius of $R_0\sim 1\mu$m, the  kick frequency $\omega_0$ 
can be obtained from  Eqs.(\ref{eq:5-2b}) as
$
\omega_0 = \frac{\hbar}{m^*R_0^2}\,\frac{1}{{\tilde\hbar}}
         \simeq \frac{2}{{\tilde\hbar}} \,{\rm GHz}.$
Then the required $B_{dc}$ and $B_{ac}$ magnetic fields have the values:
$B_{dc} = \frac{\omega_0 m^* c}{e^*}\,{{\tilde\omega_c}}
       \simeq 20 \frac{{\tilde\omega_c}}{{\tilde\hbar}}$ {\rm gauss}, and
$B_{ac} = \epsilon B_{dc} \simeq 
           20 \frac{\epsilon{\tilde\omega_c}}{{\tilde\hbar}}
\,{\rm gauss}.$  The $a.c.$ Larmor frequency is
 $\omega_{ac} = \epsilon {\tilde\omega_c} \simeq 
20 \,\frac{\epsilon{\tilde\omega_c}}{{\tilde\hbar}}\, 
{\rm MHz}.$ 
 With these values, in the quasi-integrable regime,
with parameters  $(\epsilon,{\tilde\omega_c})^{(reg)} = (0.1,0.1)$, 
we get
$
\omega_0^{(reg)} \simeq 20 \,{\rm GHz}$ and
$B_{ac}^{(reg)} \simeq 20 \,{\rm gauss}.
$
In the chaotic regime we take the parameters 
$(\epsilon,{\tilde\omega_c})^{(chaos)} = (2.0,2.0)$.
which leads to 
$\omega_0^{(chaos)} \simeq 20 \,{\rm GHz}$
and $B_{ac}^{(chaos)} \simeq 800 \,{\rm gauss}$.
These results for the regular and chaotic regimes are within 
experimental reach. 

\section*{Acknowledgments}

This work has been supported in part by CONACYT 3047P and by NSF grant 
DMR-9521845.

\newpage
\appendix
\section{}

In this appendix we provide the derivation of Eq.(\ref{eq:eq5-35}).
For any operator $\hat A$, the expectation value of the linear
response under the action of a constant Hamiltonian $H_0$ and a
time-dependent perturbation  $V(t)$ is given by
\begin{equation}
\label{eq:eq5-lr1}
\langle \hat A\rangle(t) = \langle \hat A\rangle_0
+ \frac{1}{i\hbar}\int_0^t 
dt_1 \langle[\hat A(t_1-t),\hat V(t_1)]\rangle_0,
\end{equation}
where $\langle\,\cdot\,\rangle_0 = Tr\{\rho_0\cdot\}$, and $\rho_0$ is 
the density matrix associated with the unperturbed Hamiltonian $H_0$. 
Thus,
\begin{eqnarray}
\label{alleq5-lr2}
\langle[\hat A(t_1-t),\hat V(t_1)]\rangle_0 
&=& Tr\left\{\rho_0[\hat A(t_1-t),\hat V(t_1)]\right\} \nonumber \\
&=& Tr\left\{\rho_0 e^{iH_0(t_1-t)/\hbar} \hat A e^{-iH_0(t_1-t)/\hbar}
\hat V(t_1)\right.
\nonumber \\
&-& \left. \rho_0 \hat V(t_1) e^{iH_0(t_1-t)/\hbar} 
\hat A e^{-iH_0(t_1-t)/\hbar}\right\}. 
\end{eqnarray}
For a single particle pure state $|\psi_{n_i}\rangle$, 
$\rho_0 = |\psi_{n_i}\rangle\langle\psi_{n_i}|$. Thus, from above, 
\begin{eqnarray}
\label{alleq5-lr3}
\langle\dots\rangle_0 
&=& \sum_n\langle\psi_n|\psi_{n_i}\rangle\langle\psi_{n_i}|
e^{iH_0(t_1-t)/\hbar} \hat A e^{-iH_0(t_1-t)/\hbar}V(t_1)|\psi_n\rangle
\nonumber\\
&-& \sum_n\langle\psi_n|\psi_{n_i}\rangle\langle\psi_{n_i}|
\hat V(t_1)e^{iH_0(t_1-t)/\hbar} \hat A e^{-iH_0(t_1-t)/\hbar}|
\psi_n\rangle\nonumber\\
&=& e^{iE_{n_i}(t_1-t)/\hbar}\langle\psi_{n_i}|\hat
 Ae^{-iH_0(t_1-t)/\hbar}
\hat V(t_1)|\psi_{n_i}\rangle \nonumber \\
&-& e^{-iE_{n_i}(t_1-t)/\hbar}\langle\psi_{n_i}|\hat
 V(t_1)e^{-iH_0(t_1-t)/\hbar}
\hat A|\psi_{n_i}\rangle \nonumber \\
&=& e^{iE_{n_i}(t_1-t)/\hbar}\sum_n e^{-iE_{n_i}(t_1-t)/\hbar}
\langle\psi_{n_i}|\hat A|\psi_n\rangle\langle\psi_n|\hat V(t_1)|
\psi_{n_i}\rangle\nonumber\\
&-& e^{-iE_{n_i}(t_1-t)/\hbar}\sum_n e^{iE_{n_i}(t_1-t)/\hbar}
\langle\psi_{n_i}|\hat V(t_1)|\psi_n\rangle\langle\psi_n|\hat A|
\psi_{n_i}\rangle .
\end{eqnarray}
In our case,
\begin{eqnarray}
\label{alleq5-lr4}
\hat A = \tilde M_z &=& -\frac{\hat L_z}{\tilde\hbar} - 
\frac{f}{2}\hat r^2 \nonumber \\
\hat V(t_1) &=& {\tilde V}\sum_{p=0}^{N_T}\delta(t_1-p).
\end{eqnarray}
Since the $\left\{|\psi_n\rangle\right\}$ are real and $\hat A$ and $
\hat V$  are Hermitian, 
\begin{equation}
\label{eq:5-lr5}
\langle\dots\rangle_0 = \sum_n A_{n_i,n} V_{n,n_i}(t_1)
\left(e^{i\omega_{n_i,n}(t_1-t)} - e^{-i\omega_{n_i,n}(t_1-t)}\right),
\end{equation}
where
\begin{eqnarray}
\label{alleq5-lr6}
A_{n_i,n} &=& -\ell_i\delta_{n_i,n} - 
\frac{f}{2}\langle\psi_{n_i}|r^2|\psi_n\rangle 
\equiv -\ell_i\delta_{n_i,n} - \frac{f}{2} \langle r^2\rangle_{n_i,n}, \\
V_{n,n_i} &=& \frac{1}{2}\left(\frac{\epsilon{\tilde\omega_c}}{2}\right)^2 
\langle r^2\rangle_{n,n_i}\sum_p\delta(t_1-p),\quad{\rm and} \\
\omega_{n_i,n} &=& \frac{E_{n_i}-E_n}{\tilde\hbar} .
\end{eqnarray}
Finally, we have 
\begin{eqnarray}
\label{eq:eq5-lr7}
\langle\dots\rangle_0 = -i\left(\frac{\epsilon{\tilde\omega_c}}
{2}\right)^2 
&\sum_n& \langle r^2\rangle_{n,n_i}\left(\ell_i\delta_{n_i,n}
+ \frac{f}{2}  \langle r^2\rangle_{n,n_i}\right) \nonumber \\
&\sum_{p=0}^{N_T}&
\sin\left\{\omega_{n_i,n}(t_1-t)\right\}\delta(t_1-p)
\end{eqnarray}
and, 
\begin{equation}
\label{eq:eq5-lr8}
\langle A\rangle_0 = \langle\psi_{n_i}|\tilde M_z|\psi_{n_i}\rangle = 
-\ell_i - \frac{f}{2}\langle r^2\rangle_{n_i,n_i}.
\end{equation} 
Substituting Eqs.(\ref{eq:eq5-lr7}) and (\ref{eq:eq5-lr8}) into 
Eq.(\ref{eq:eq5-lr1}), we  get Eq.(\ref{eq:eq5-35}). This completes our 
derivation of the linear response result. 

\newpage

%
%

%
\newpage
%

\vskip 1cm
\begin{figure}[hp]\begin{center}
\caption{ Classical phase diagram separating the regular
from chaotic regions, used in the analysis of this paper. 
See text for the definition of the dimensionles variables in the axes.
\label{fig1}}
\end{center}\end{figure}

\begin{figure}[hp]\begin{center}
\caption{(a) Single electron time-dependent magnetization 
$\langle M\rangle(t)$ for $\epsilon=0.1$, $\tilde{\omega_c}=0.1$, 
and $\tilde\hbar=0.1$ in the quasi-regular regime
(see Fig. \ref{fig1}). 
(b) Power Spectrum, $S(\nu$), of $\langle M\rangle$ corresponding to 
the parameter values of Fig. \ref{fig2}(a). The scale of $\nu$ frequencies 
is arbitrary. In all figures the axis are given in terms of
dimensionless units. 
\label{fig2}}
\end{center}\end{figure}

\begin{figure}[hp]\begin{center}
\caption{(a) Same as in Fig. \ref{fig2}(a) for 
$\epsilon=1.0$, $\tilde {\omega_c}=1.0$, $\tilde\hbar=0.1$, 
which is in the regime
that is approaching the chaotic region in parameter space
(see Fig. \ref{fig1}). 
(b) Same as in Fig. \ref{fig2}(b) for the 
parameters of Fig. \ref{fig3}(a).
\label{fig3}}
\end{center}\end{figure}

\begin{figure}[hp]\begin{center}
\caption{(a) The magnetization as a function of time for the parameters 
$\epsilon=2.0$, $\tilde{\omega_c}=2.0$, $\tilde\hbar=0.1$, deep in the chaotic
regime (see Fig. \ref{fig1}).  (b) Here the classical chaos is clearly 
revealed in the noisy structure 
of $<M>$ and the broad band spectrum of $S(\nu$).
\label{fig4}}
\end{center}\end{figure}

\begin{figure}[hp]\begin{center}
\caption{(a) Time-averaged magnetization $\langle\langle M\rangle\rangle$ 
(per electron) in the quasi-integrable regime for  $\epsilon=0.1$, 
$\tilde{\omega_c}=0.1$,  $\tilde\hbar=0.1$, as a function of electron  
number $N$. We see that  $\langle\langle M\rangle\rangle$ oscillates
non monotonically with $N$. See text for 
a theoretical perturbative explanation of these oscillations. 
(b) Total time-averaged orbital magnetization 
(per electron) as a function of $N$ for the same 
parameter values as in Fig. \ref{fig5}(a). Here we see, as expected,
a direct correspondence with the magnetization of Fig. \ref{fig5}(a).
(c) Same as in Fig. \ref{fig5}(a) for parameters in the chaotic regime,
$\epsilon=2.0$, $\tilde{\omega_c}=2.0$, $\tilde\hbar=0.1$. 
In this case $\langle\langle M\rangle\rangle$
oscillates but its behavior is {\it purely} diamagnetic.
(d) Orbital current for the chaotic parameter values of
Fig. \ref{fig5}(c), with a clear correspondence to the purely diamagnetic
nature of the magnetic response of the system in the chaotic regime.
\label{fig5}}
\end{center}\end{figure}

\begin{figure}[hp]\begin{center}
\caption{(a) Time-averaged magnetization for one electron as a function of
${\tilde\omega_c}$ for $\epsilon=0.1$, and ${\tilde\hbar}=0.1$, with 
increment in
diamagnetism as ${\tilde\omega_c}$ increases.
(b) Time-averaged energy as a function of electron number, with 
a clear quadratic
dependence, for the same parameters as in (a).
(c)  $\langle\langle M\rangle\rangle$ 
in the chaotic regime for one electron, with $\epsilon=10$, 
${\tilde\hbar}=0.1$, as a function
of ${\tilde\omega_c}$. Here we also note an increase of 
diamagnetism as ${\tilde\omega_c}$
grows.
(d) $\langle\langle E\rangle\rangle$ for the same parameters as 
in (b) in the chaotic regime with a 
clear linear dependence
on $N$. See text for further discussion of these results.
\label{fig6}}
\end{center}\end{figure}

\begin{figure}[hp]\begin{center}
\caption{Comparison between exact ($\Box$) and perturbative ($+$) calculations 
for the time-averaged magnetization as a function of electron number. 
The parameters here are, $\epsilon=0.01$, 
${\tilde\hbar}=0.1$, and ${\tilde\omega_c}=0.01$. 
Note the almost exact agreement 
between the two calculations, with the magnitude of the diamagnetic 
response much smaller than that of the paramagnetic response. 
\label{fig7}}
\end{center}\end{figure}


\end{document}